\documentclass[11pt]{article}
\usepackage{moriond,epsfig}

\def\Journal#1#2#3#4{{#1} {\bf #2}, #3 (#4)}

\def\NPB{{\em Nucl. Phys.} B}

\def\be{\begin{equation}}
\def\ee{\end{equation}}
\def\bea{\begin{eqnarray}}
\def\eea{\end{eqnarray}}

\begin{document}
\begin{flushright}
  Cavendish-HEP-05/07
\end{flushright}
\vspace*{4cm}
\title{CP VIOLATION IN CHARGED HIGGS PRODUCTION}

\author{ J.A. WILLIAMS }

\address{High Energy Physics Group, Cavendish Laboratory, University of Cambridge,
  Madingley Road, Cambridge, CB3 0HE, United Kingdom}

\maketitle\abstracts{I present a study estimating the amount of CP violation present in
  the production of charged Higgs bosons due to the presence of complex
  trilinear scalar couplings.  I compare the results of my study with previous results for
  the decay.  I briefly comment on the possibility of observing this CP asymmetry at the
  LHC.}

I have investigated the possibility of observing CP violation in the production and decay
of MSSM charged Higgs bosons at the LHC.  The CP violation arises from allowing the
trilinear scalar couplings in the soft breaking Lagrangian to be complex, leading to
complex phases.  For the initial study I have chosen to investigate the effect of a
complex $A_t$, keeping the other phases zero.

CP violation can occur at the $H^\pm t b$ vertex due to vertex corrections and Higgs
self-energy diagrams containing supersymmetric particles.
The decay process $H^\pm \to t b$ has already been studied~\cite{Christova:2002ke}.  I
have investigated the production with a view to combining both the production and the
decay to obtain a complete description.  The amount of CP violation is given by the asymmetry
\begin{equation}
  \label{eq:1}
  \mathcal{A}_{\mathrm {CP}} = \frac{\sigma(H^+) - \sigma(H^-)}{\sigma(H^+) + \sigma(H^-)}.
\end{equation}

\section{The Decay}
\label{sec:decay}

Initially I tried to reproduce the results for the decay to ensure that the production and
decay studies would be consistent.  While the sub-dominant loops agreed (such as the loops
containing $\tilde t \tilde b \tilde \chi^0$ and $\tilde \chi^0 \tilde \chi^\pm \tilde t /
\tilde b$) there was a discrepancy in the dominant loop containing $\tilde t \tilde b
\tilde g$.  I discovered that this was due to a conjugation error in
ref~\cite{Christova:2002ke}.  After correcting this, the two methods for calculating the
decay asymmetry agree.  The corrected asymmetry is less than that in the original study.

\section{The Production}
\label{sec:production}

There are two main processes for charged Higgs production at the LHC, bottom -- gluon
fusion and gluon -- gluon fusion.  Care needs to be taken when combining them to avoid
double counting.  Bottom -- gluon fusion is the dominant process and is the only one
considered in this initial study.

\begin{figure}[htbp]
  \centering
  \begin{tabular}{@{}c @{} c @{}}
  \hspace{-0.25cm}\resizebox{8.5cm}{!}{\includegraphics{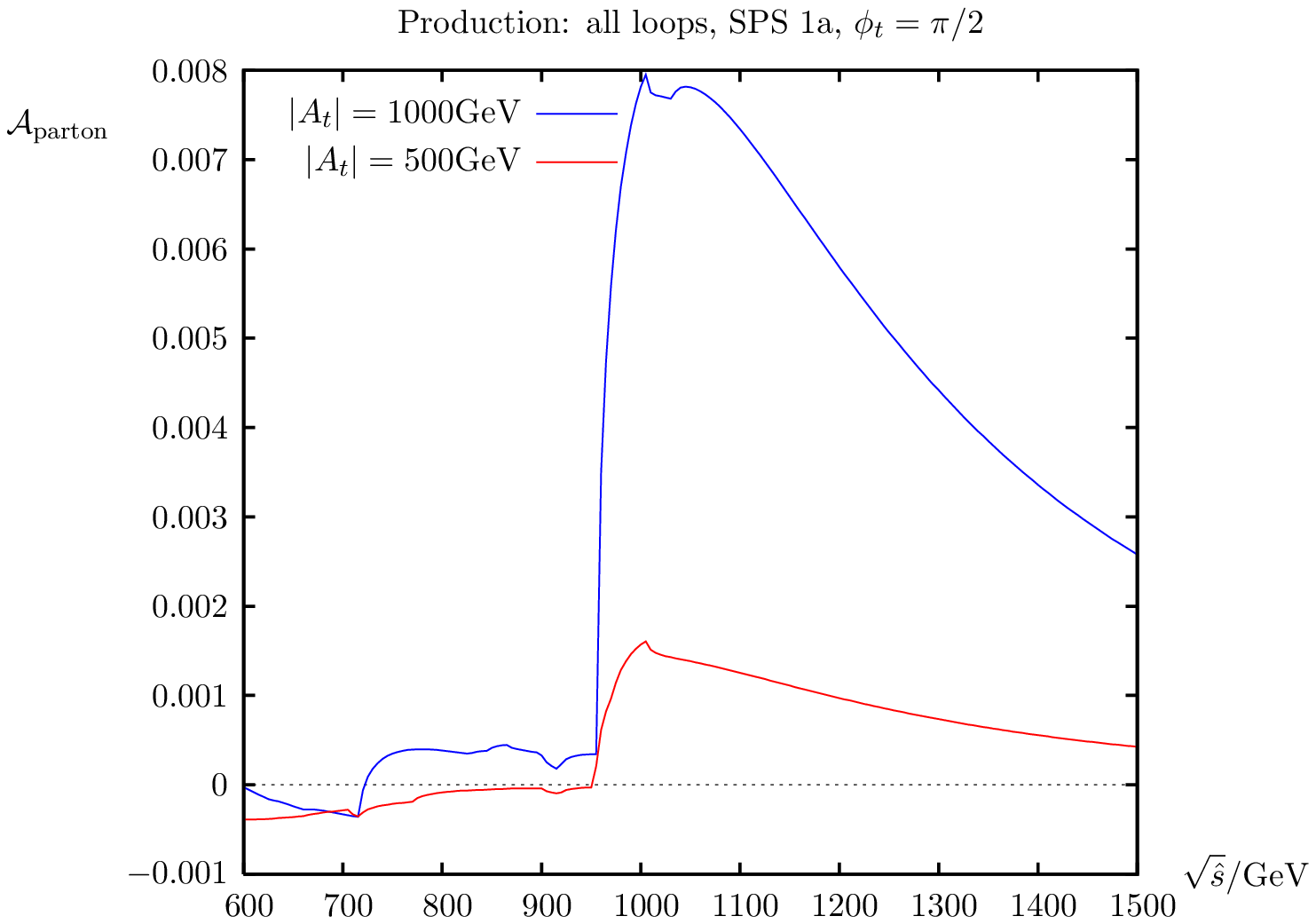}}
  &
  \hspace{-0.5cm}\resizebox{8.5cm}{!}{\includegraphics{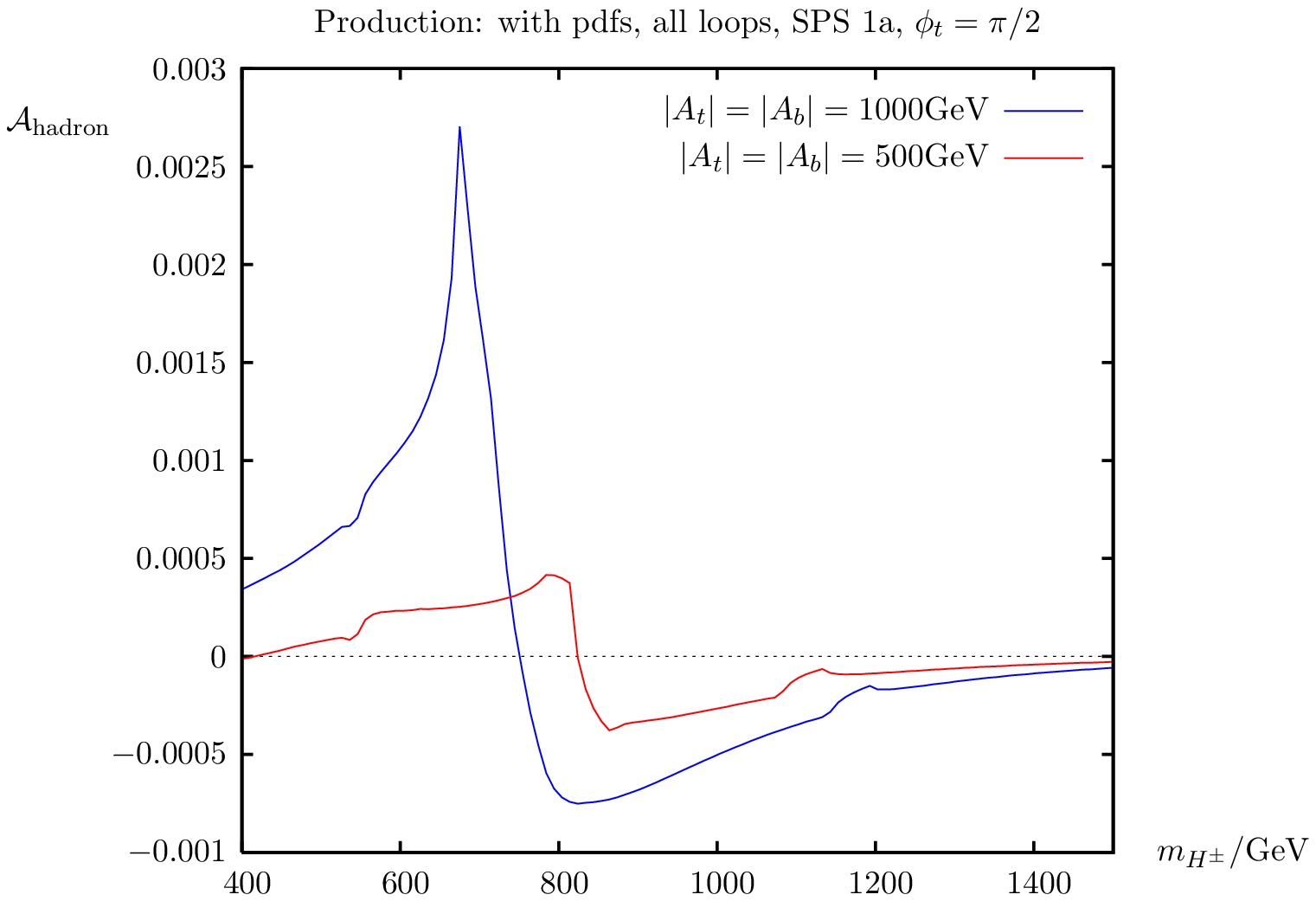}}
  \end{tabular}
  \caption{CP asymmetry in charged Higgs production (a) parton level.  (b) hadron level.}
  \label{fig:asymmetry}
\end{figure}

The cross-section was calculated using {\textsf {FormCalc}}~\cite{Hahn:1998yk}.  The
parton level results are shown in Figure~\ref{fig:asymmetry}(a).  The thresholds in
partonic centre of mass energy, $\sqrt{\hat s}$, appear when different particles in the
vertex correction can be produced as real particles, giving rise an imaginary part of the
amplitude, which allows CP violation to be manifest.

Since the LHC is a proton -- proton collider, it is necessary to include the
parton distribution functions (pdfs) for the bottom quarks and gluons to obtain the CP
asymmetry at the hadron level, Figure~\ref{fig:asymmetry}(b).  Including
the pdfs greatly reduces the asymmetry, making it very challenging to observe at the LHC.

\section{Observing the Asymmetry at the LHC}
\label{sec:observ-asymm}

The total cross-section is approximately $50\:$pb for low charged Higgs mass.  For an
integrated luminosity at the LHC of $100\:$fb$^{-1}$, this would give a total of $5\,000$
charged Higgs events.  Reducing this to take account of an acceptance of, say, 5\%, gives
only $250$ events.  It would not be possible to observe an asymmetry as small as that in
Figure~\ref{fig:asymmetry} with so few events and it is likely that it will be necessary
to wait for the luminosity upgrade of the LHC to be able to observe this asymmetry.

Work which is being done to extend this initial study involves combining the production and
decay asymmetries; including the gluon -- gluon fusion production process; investigating
other points in the MSSM parameter space; varying other parameters such as the phase of
the trilinear couplings; investigating other loop processes such as box diagrams and
exploring the possibilities of seeing the asymmetry at a future International Linear
Collider which would provide a cleaner environment.  This work forms part
of ref~\cite{Williams}.

\end{document}